# Design Principles for Sparse Matrix Multiplication on the GPU

Carl Yang[1,2],[0000−0002−4357−0906], Aydın Buluç[2,3],[0000−0001−7253−9038]
, and John D. Owens[1,2][0000−0001−6582−8237]

[1] University of California, Davis CA 95616, USA
[2] Lawrence Berkeley National Laboratory, Berkeley CA 94720, USA
[3] University of California, Berkeley CA 94720, USA

**Abstract.** We implement two novel algorithms for sparse-matrix dense-matrix multiplication (SpMM) on the GPU. Our algorithms expect the sparse input in the popular compressed-sparse-row (CSR) format and thus do not require expensive format conversion. While previous SpMM work concentrates on thread-level parallelism, we additionally focus on latency hiding with instruction-level parallelism and load-balancing. We show, both theoretically and experimentally, that the proposed SpMM is a better fit for the GPU than previous approaches. We identify a key memory access pattern that allows efficient access into both input and output matrices that is crucial to getting excellent performance on SpMM. By combining these two ingredients—(i) merge-based load-balancing and (ii) row-major coalesced memory access—we demonstrate a 4.1× peak speedup and a 31.7% geomean speedup over state-of-the-art SpMM implementations on real-world datasets.

## 1 Introduction

Many algorithms in machine learning, data analysis, and graph analysis can be organized such that the bulk of the computation is structured as sparse matrix-dense matrix multiplication (SpMM). Examples include inference on pruned neural networks [1], graph centrality calculations [2], all-pairs shortest paths [3], iterative solvers with multiple righthand sides [4], blocked eigensolvers such as Blocked Lanczos [5] or Locally Optimal Block Preconditioned Conjugate Gradient (LOBPCG) [6], sparse matrix precision estimation [7], multi-scale spectral graph decomposition [8], non-negative matrix factorization [9], and tomographic reconstruction [10]. SpMM is also one of the possible instantiations of the most prevalent GraphBLAS primitive, namely the matrix-matrix multiplication operation on a semiring (GrB_mxm) [11], depending on the sparsity of operands.

Given an $m$-by-$k$ sparse matrix $\mathbf{A}$ and a $k$-by-$n$ dense matrix $\mathbf{B}$, SpMM computes an $m$-by-$n$ dense matrix $\mathbf{C} = \mathbf{AB}$. We assume $n \ll m$ and $n \gg k$, that is to say, SpMM is multiplying a sparse matrix with a tall-skinny dense matrix. We choose the most common sparse matrix format—compressed sparse row (CSR)—because we avoid the substantial cost of matrix conversion. However, CSR results in a challenging problem on the GPU, because the sparse row can vary significantly in how many nonzeroes there are. We combine recent advances from the related problem of sparse matrix-dense vector multiplication (SpMV) [12–14] and a key memory access pattern we identify as critical to SpMM performance in order to propose and implement two SpMM algorithms that demonstrate superior performance to state-of-the-art specialized matrix formats and vendor-supplied CSR SpMM implementations.

Our main contributions in this paper are:

1. We generalize two main classes of SpMV algorithms—(1) row splitting and (2) merge-based—for the SpMM problem and implement them on the GPU.
2. We introduce a simple heuristic that selects between the two kernels with an accuracy of 99.3% compared to optimal.
3. Using our multi-algorithm and heuristic, we achieve a geomean speed-up of 31.7% and up to a maximum of 4.1x speed-up over state-of-the-art SpMM implementations over 157 datasets from the SuiteSparse Matrix Collection [15].

## 2 Background & Preliminaries

### 2.1 GPUs

Modern GPUs are throughput-oriented manycore processors that rely on large-scale multithreading to attain high computational throughput and hide memory access time. The latest generation of NVIDIA GPUs have up to 80 "streaming multiprocessors" (SMs), each with up to hundreds of arithmetic logic units (ALUs). GPU programs are called *kernels*, which run a large number of threads in parallel in a single-program, multiple-data (SPMD) fashion.

The underlying hardware runs an instruction on each SM on each clock cycle on a *warp* of 32 threads in lockstep. The largest parallel unit that can be synchronized within a GPU kernel is called a *cooperative thread array* (CTA), which is composed of warps. For problems that require irregular data access, a successful GPU implementation needs to (1) ensure coalesced memory access to external memory and efficiently use the memory hierarchy, (2) minimize thread *divergence* within a warp, and (3) maintain high *occupancy*, which is a measure of how many threads are available to run on the implementation on the GPU.

### 2.2 Sparse matrix formats and SpMM

An $m \times n$ matrix is often called *sparse* if its number of nonzeroes $nnz$ is small enough compared to $\mathcal{O}(mn)$ such that it makes sense to take advantage of sparsity. The compressed sparse row (CSR) format stores only the *column indices* and *values* of nonzeroes within a row. The start and end of each row are then stored in terms of the column indices and value in a *row offsets* (or *row pointers*) array. Hence, CSR only requires $m + 2nnz$ memory for storage. We say a dense matrix is in *row-major order* when successive elements in the same row are contiguous in memory. Similarly, we say it is in *column-major order* when successive elements in the same column are contiguous in memory.

Similarly to sparse matrix-dense vector multiplication (SpMV), a desire to achieve good performance on SpMM has inspired innovation in matrix storage formatting [16–18]. These custom formats and encodings take advantage of the matrix structure and underlying machine architecture. Even only counting GPU processors, there exist more than sixty specialized SpMV algorithms and sparse matrix formats [19].

The vendor-shipped library cuSPARSE library provides two functions csrmm and csrmm2 for SpMM on CSR-format input matrices [20]. The former expects a column-major input dense matrix and generates column-major output, while the latter expects row-major input and generates column-major output. Among many efforts to define and characterize alternate matrix formats for SpMM are a variant of ELLPACK called ELLPACK-R [16] and a variant of Sliced ELLPACK called SELL-P [17]. Hong et al. performs dynamic load-balancing by separating the sparse matrix into heavy and light rows. The heavy rows are processed by CSR and the light rows by doubly compressed sparse row (DCSR) in order to take advantage of tiling [21].

However, there is a real cost to deviating from the standard CSR encoding. Firstly, the rest of the computation pipeline will need to convert from CSR to another format to run SpMM and convert back. This process may take longer than the SpMM operation itself. Secondly, the pipeline will need to reserve valuable memory to store multiple copies of the same matrix—one in CSR format, another in the format used for SpMM.

## 3 Design Principles

In this section, we discuss two design principles that every irregular problem on the GPU must follow for good performance. Ideally, we attain *full utilization* of the GPU hardware, where a ready warp can be run on every cycle, all computational units are doing useful work on every cycle, and all memory accesses are coalesced. Our principles for reaching this goal are (1) effective latency-hiding through a combination of thread- and instruction-level parallelism (TLP and ILP) and (2) efficient load-balancing. Then we will look at state-of-the-art SpMM implementations to understand their inefficiencies.

### 3.1 Latency hiding with TLP and ILP

Memory operations to a GPU's main memory take hundreds of clock cycles. The GPU's primary technique for hiding the cost of these long-latency operations is through thread-level parallelism (TLP). Effective use of TLP requires that the programmer give the GPU enough work so that when a GPU warp of threads issues a memory request, the GPU scheduler puts that warp to sleep and another ready warp becomes active. If enough warps are resident on the GPU (if we have enough TLP), switching between warps can completely hide the cost of a long-latency operation. We quantify the amount of TLP in a program as *occupancy*, the ratio of available (issued) warps to the maximum number of warps that can be supported by the GPU. Higher occupancy yields better latency-hiding ability, which allows us to approach full utilization.

Another latency-hiding strategy is exploiting instruction-level parallelism (ILP) and its ability to take advantage of overlapping the latency of multiple memory operations within a single thread. Because the GPU's memory system is deeply pipelined, a thread can potentially issue multiple independent long-latency operations before becoming inactive, and those multiple operations will collectively incur roughly the same latency as a single operation. While this yields a significant performance advantage, it relies on the programmer exposing independent memory operations to the hardware. We can achieve this goal by assigning multiple independent tasks to the same thread ("thread coarsening").

GPUs have a fixed number of registers. TLP requires many resident warps, each of which requires registers. ILP increases the work per thread, so each thread requires more registers. Thus TLP and ILP are in opposition, and attaining full utilization requires carefully balancing both techniques. While TLP is commonly used across all of GPU computing, ILP is a less explored area, with prior work limited to dense linear algebra [22] and microcode optimization [23].

### 3.2 Load-balancing

We now turn to the problem of ensuring that all computational units are doing useful work on every cycle, and that the memory accesses from those warps are coalesced to ensure peak memory performance. In the context of SpMV and SpMM, this "load-balancing" problem has two aspects:

1. Load imbalance *across* warps. Some CTAs or warps may be assigned less work than others, which may lead to these less-loaded computation units being idle while the more loaded ones continue to do useful work. In this paper, we term this "Type 1" load imbalance.
2. Load imbalance *within* a warp, in two ways, which we collectively call "Type 2" load imbalance. (a) Some warps may not have enough work to occupy all 32 threads in the warp. In this case, thread processors are idle, and we lose performance. (b) Some warps may assign different tasks to different threads. In this case, SIMD execution within a thread means that some threads are idle while other threads are running; moreover, the divergence in execution across the warp means memory accesses across the entire warp are unlikely to be coalesced.

For irregular matrices, we claim that SpMV and SpMM are fundamentally load-balancing problems on the GPU. As evidence, Figure 1 shows load imbalance in a vendor-supplied implementation from a synthetic benchmark. The experimental setup is described in Section 5. The right side of the $x$-axis represents Type 1 load imbalance, where long matrix rows are not divided enough, resulting in some computation resources on the GPU remaining idle while others are overburdened. The left size of the $x$-axis represents Type 2 load imbalance where too many computational resources are allocated to each row, so some remain idle.

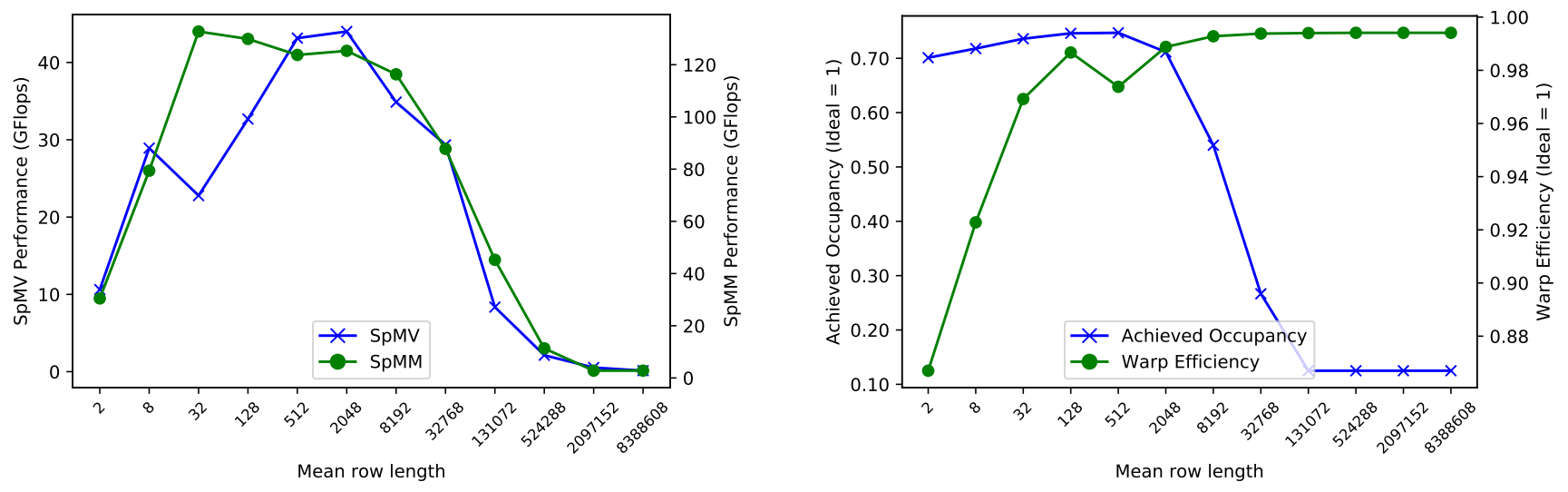

(a) SpMV and SpMM (b) Occupancy and warp efficiency

**Fig. 1.** Synthetic benchmark showing NVIDIA cuSPARSE SpMV and SpMM performance as a function of matrix dimensions on a Tesla K40c, and SpMM's achieved occupancy and warp efficiency (inverse of divergence).

## 4 Parallelizations of CSR SpMM

This section reviews three existing parallelizations of SpMV through the lens of the design principles from Section 3. While our implementations of SpMM share some characteristics with SpMV parallelizations, we also faced several different design decisions for SpMM, which we discuss below. The three SpMV variants are illustrated in Figure 2 and summarized here:

1. Row split [24]: Assigns an equal number of rows to each processor.
2. Merge based: Performs two-phase decomposition—the first kernel divides work evenly amongst CTAs, then the second kernel processes the work.
   (a) Nonzero split [12,13]: Assign an equal number of nonzeroes per processor. Then do a 1-D (1-dimensional) binary search on *row offsets* to determine at which row to start.
   (b) Merge path [14]: Assign an equal number of {nonzeroes and rows} per processor. This is done by doing a 2-D binary search (i.e., on the diagonal line in Figure 2(c)) over *row offsets* and *nonzero indices* of matrix $\mathbf{A}$.

While row split focuses primarily on ILP and TLP, nonzero split and merge path focus on load-balancing as well. We consider nonzero split and merge path to be *explicit load-balancing* methods, because they rearrange the distribution of work such that each thread must perform $T$ independent instructions; if $T > 1$, then explicit load-balancing creates ILP where there was previously little or none. Thus load-balance is closely linked with ILP, because if each thread is guaranteed $T > 1$ units of independent work (ILP), then each thread is doing the same amount of work (i.e., is load-balanced).

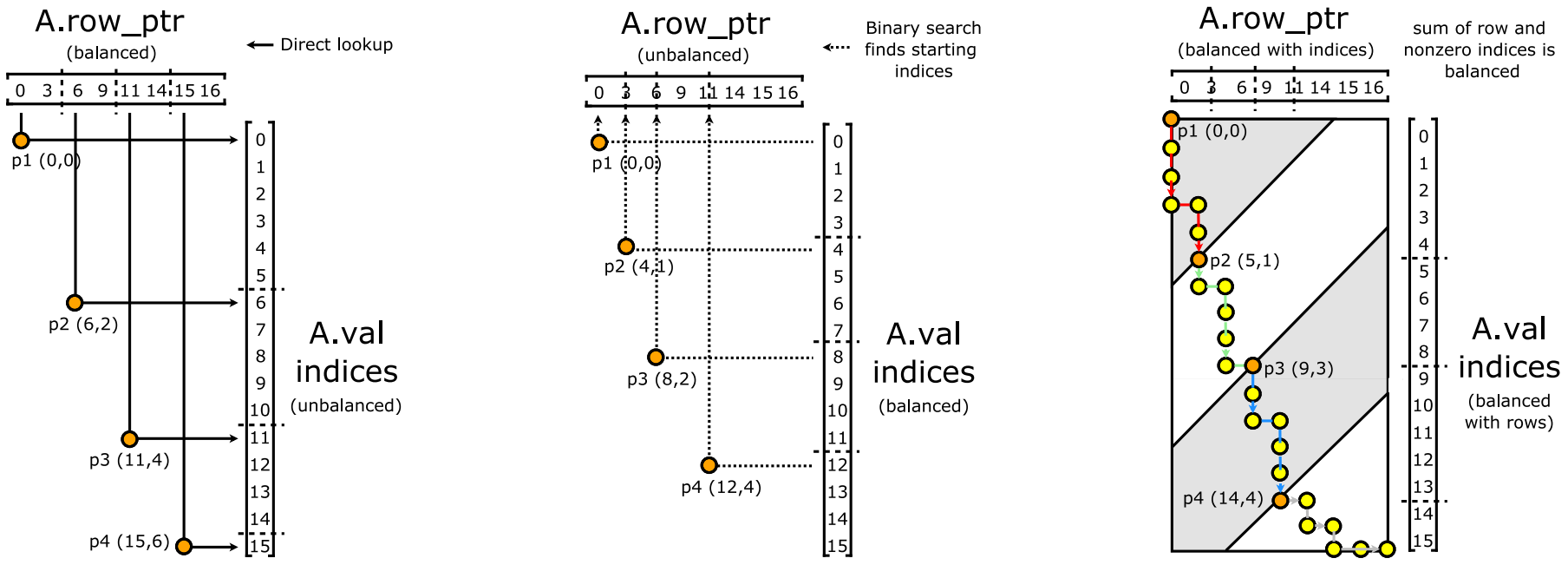

(a) Row split (b) Nonzero split (c) Merge path

**Fig. 2.** The three parallelizations for CSR SpMV and SpMM on matrix **A**. The orange markers indicate segment start for each processor ($P = 4$).

We contend that nonzero split and merge path despite having different structure possess similar performance characteristics. The binary search being done in 2-D (i.e. on the diagonal line in Figure 2(c)) as opposed to 1-D is equivalent to making an implicit assumption that a write to **C** has the same cost as a memory read from **A** and **B**. As Merrill and Garland point out, this solves the pathological case of matrices that have infinitely many empty rows. However, the merge path is more challenging to implement, so we decide to extend the Baxter's nonzero split concept [12] to SpMM under the moniker "merge-based SpMM".

### 4.1 Algorithm I: Row-splitting SpMM

Row split aims to assign each row to a different thread, warp, or CTA. Figure 3(a) shows the warp assignment version. The typical SpMV row split is only the left-most column of matrix **B** with orange cells replaced by green cells. This gives SpMV 1 independent instruction and uncoalesced, random accesses into the vector. Although row-split is a well-known method for SpMV [24], we encountered three important design decisions when extending it to SpMM:

1. Granularity: Should each row be assigned to a thread, warp, or CTA?
2. Memory access pattern: How should work be divided in fetching **B**? What is the impact on ILP and TLP?
3. Shared memory: Can shared memory be used for performance gain?

*1. Granularity.* We assigned each row to a warp compared to the alternatives of assigning a thread and a CTA per row. This leads to the simplest design out of the three options, since it gives us coalesced memory accesses into **B**. For matrices with few nonzeroes per row, the thread-per-matrix-row work assignment may be more efficient. This is borne out by Figure 4.

*2. Memory access pattern.* This design decision had the greatest impact on performance. To our knowledge, this is the first time in literature this novel memory access strategy has been described. Our thread layout is shown in Figure 3(c). For SpMM, we have two approaches we could take: each thread is responsible for loading a column or a row of the matrix **B**.

We discovered the first approach is better, because the memory accesses into **B** are independent and can be done in a coalesced manner (provided that **B** is in row-major order). In contrast, memory accesses into a column-major **B** would be independent but uncoalesced. Compared to the SpMV case, each thread now has 32 independent instructions and coalesced memory accesses into **B**, which significantly amortizes the cost of memory accesses compared to accessing a single vector. However, since we are forcing threads to pass a dummy column index if they are out of bounds within a row, the effective number of independent instructions and coalesced memory accesses is sensitive to row lengths that do not divide 32. For example, if the row length is 33, then we will be doing 64 independent instructions and coalesced memory accesses into **B**. Whether or not they divide 32 does not matter for very long rows, because the cost is amortized by efficiently processing batches of 32. However, we would expect row split to

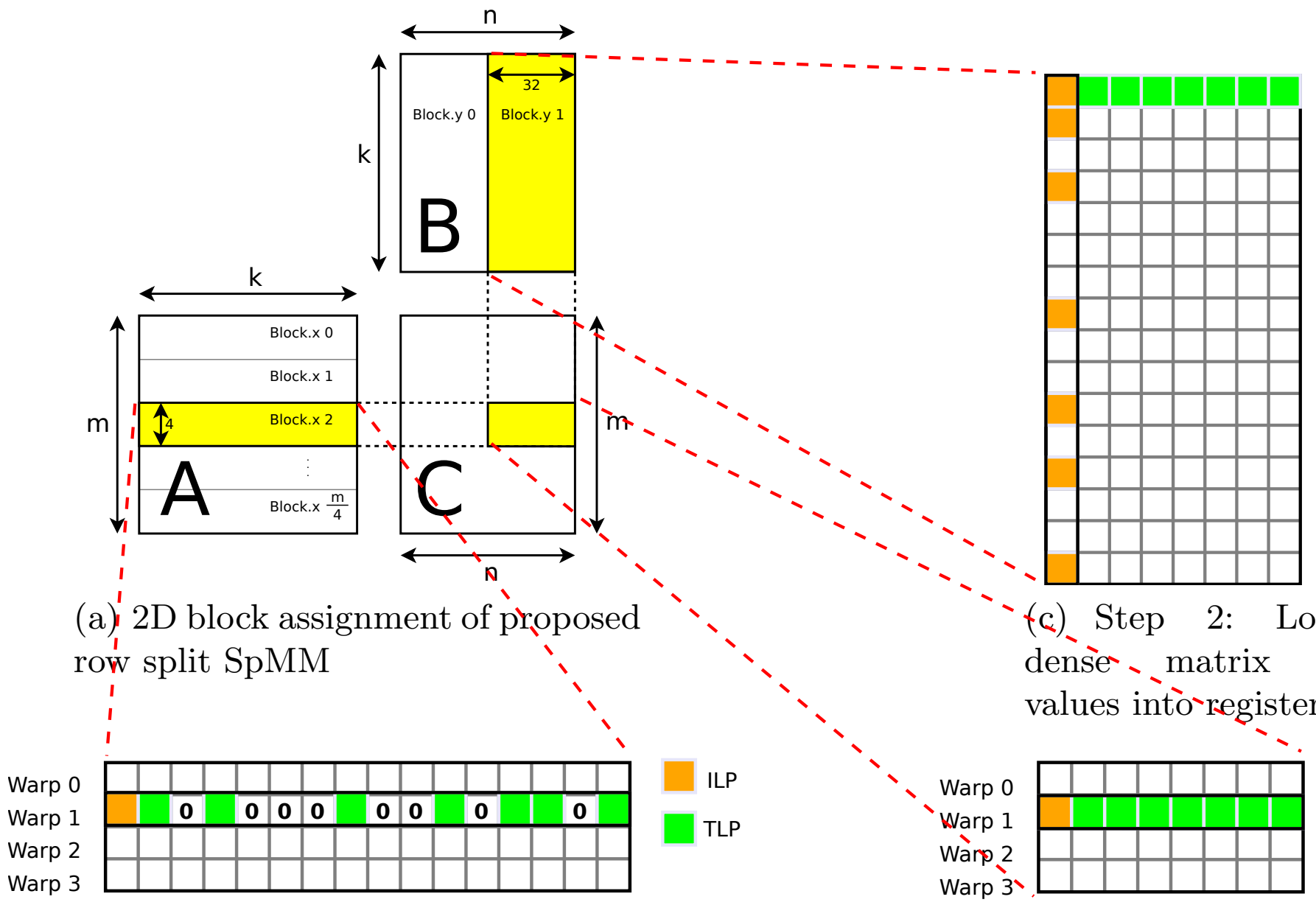

(a) 2D block assignment of proposed row split SpMM

(c) Step 2: Load dense matrix **B** values into registers.

(b) Step 1: Load sparse matrix **A** column indices into registers. Zeroes are marked as such.

(d) Step 3: Write to dense matrix **C**.

**Fig. 3.** Figure 3(a) shows the tiling scheme we use. Figures (b), (c), (d) represent the yellow blocks from Figure 3(a). Row split SpMM ILP (orange) and TLP (green) are shown using warp 1 with 8 threads per warp. In practice, we use 32 threads per warp and 4 warps per GPU cooperative thread array (CTA). Matrix **A** is sparse in CSR format. Matrices **B** and **C** are both dense in row-major format.

be negatively impacted by Type 2 load imbalances. The summary of this ILP analysis is shown in Table 1.

**Table 1.** This table shows the number of independent instructions per GPU thread for SpMV and SpMM with default value shown in brackets, as well as the register usage and the extra number of memory accesses with respect to the row-split algorithm. $T$ is the number of work items per thread (typically specified as a tuning parameter to the algorithm). $L$ is the number of nonzeroes modulus 32 in the row of $\mathbf{A}$ that we are computing. $B$ is the CTA size. Typical values for $T$ in SpMV and SpMM are 7 and 1 respectively, while a typical value for $B$ is 128. $T$ cannot be set arbitrarily high, because high register usage causes lower occupancy. $\mathbf{A}.nnz$ is the number of nonzeroes in the sparse matrix $\mathbf{A}$. $\mathbf{B}.ncols$ is the number of columns of the dense matrix $\mathbf{B}$.

| | SpMV | | SpMM | |
|---|---|---|---|---|
| Operation | Row-split | Merge-based | Row-split | Merge-based |
| Read **A**.col_ind and **A**.val | 1 | $T(7)$ | 1 | $T(1)$ |
| Read **x** / Read **B** | 1 | $T(7)$ | $0 < L \leq 32$ | $32T(32)$ |
| Write **y** / Write **C** | 1 | $T(7)$ | 1 | $32T(32)$ |
| Register usage | 2 | $2T(14)$ | 64 | $64T(64)$ |
| Memory access overhead | 0 | $\frac{\mathbf{A}.nnz}{B \times T}$ $(\frac{\mathbf{A}.nnz}{896})$ | 0 | $\frac{\mathbf{B}.ncols \times \mathbf{A}.nnz}{B \times T}$ $(2\mathbf{A}.nnz)$ |

*3. Shared memory.* The key component required is a round of 32 broadcasts (using the "shuffle" warp intrinsic __shfl) by each thread to inform all other threads in the warp which **B** row ought to be collectively loaded by the entire warp. This is required or otherwise each thread would be responsible for loading its own row, which would result in uncoalesced access. We could have also implemented this using shared memory, but since all threads are within a single warp, there is no disadvantage to preferring warp intrinsics. That they are within a single warp is a consequence of our decision to assign each row to a warp rather than a CTA.

### 4.2 Algorithm II: Merge-based SpMM

The essence of merge-based algorithms is to explicitly and evenly distribute the nonzeroes across parallel processors. It does so by doing a two-phase decomposition: In the first phase (PartitionSpmm), it divides the work between threads so that $T$ work is assigned per thread, and based on this assignment deduces the starting indices of each CTA. Once coordinated thusly, work is done in the second phase. In theory, this approach should eliminate both Type 1 and Type 2 load imbalances, and performs well in recent SpMV implementations [14]. We made the following design decisions when generalizing this technique to SpMM:

*1. Memory access pattern.* For fetching **B**, we adapt the memory access pattern that was successful in row-splitting. However, here, we must first apply the first phase (i.e., PartitionSpmm, Line 2 of Algorithm 1) to tell us the rows each CTA ought to look at if we want an equal number of nonzeroes per CTA. Then, we can apply the broadcast technique to retrieve **B** values using coalesced accesses.

**Algorithm 1** The merge-based SpMM algorithm.

**Input:** Sparse matrix in CSR $\mathbf{A} \in \mathbb{R}^{m \times k}$ and dense matrix $\mathbf{B} \in \mathbb{R}^{k \times n}$.
**Output:** $\mathbf{C} \in \mathbb{R}^{m \times n}$ such that $\mathbf{C} \leftarrow \mathbf{AB}$.

1: **procedure** SpmmMerge($\mathbf{A}, \mathbf{B}$)
2: limits[] ← PartitionSpmm(**A**, blockDim.x) ▷ **Phase 1:** Divide work and run binary-search
3: **for** each CTA $i$ **in parallel do** ▷ **Phase 2:** Do computation
4: num_rows ← limits[$i+1$] − limits[$i$]
5: shared.csr ← GlobalToShared(**A**.row_ptr + limits[$i$], num_rows) ▷ Read **A** and store to shared memory
6: end ← min(blockDim.x, **A**.nnz - blockIdx.x × blockDim.x)
7: **if** row_ind < end **then**
8: col_ind ← **A**.col_ind[row_ind] ▷ Read **A** if matrix not finished
9: valA ← **A**.values[row_ind]
10: **else**
11: col_ind ← 0 ▷ Otherwise do nothing
12: valA ← 0
13: **end if**
14: **for** each thread $j$ **in parallel do**
15: **for** $j = 0, 1, \ldots, 31$ **do** ▷ Unroll this loop
16: new_ind[$j$] ← Broadcast(col_ind, $j$) ▷ Each thread broadcasts
17: new_val[$j$] ← Broadcast(valA, $j$) ▷ col_ind and valA
18: valB[$j$] ← **B**[col_ind][$j$] × new_val[$j$] ▷ Read **B**
19: **end for**
20: **end for**
21: terms ← PrepareSpmm(shared.csr) ▷ Flatten CSR-to-COO
22: carryout[$i$] ← ReduceToGlobalSpmm(**C**, valB, valB) ▷ Compute partial of **C** and save carry-outs
23: **end for**
24: FixCarryout(**C**, limits, carryout) ▷ Carry-out fix-up (rows spanning across blocks)
25: **return C**
26: **end procedure**

*2. Register usage.* Since we opted for the coalesced memory access pattern explained in the row-splitting section, we require 32× the number of registers in order to store the values. Due to this limitation, the number of independent instructions per thread $T$ is limited to 1, so we see no further latency-hiding gain from ILP over that of row-split.

*3. Memory access overhead.* There are two sources of memory access overhead compared to the row-splitting algorithm: (1) the additional GPU kernel that determines the starting rows for each block (Line 2), and (2) the write of the carry-out to global memory for matrix rows of **C** that cross CTA boundaries (Line 24). Since the user is unable to synchronize CTAs in CUDA, this is the only way the user can pass information from one CTA to another. The first source of additional memory accesses is less of a problem for SpMM compared to SpMV, because they are amortized by the increased work. The second source, however, scales with the number of **B** columns. Thus we face a trade-off between having more efficient memory access pattern (assign 32 columns per CTA so memory access is coalesced), and having less memory access overhead (assign 4 columns per CTA so $T$ can be set higher resulting in fewer CTA boundaries that need to be crossed). The first approach resulted in better performance.

## 5 Experimental Results

### 5.1 Experimental Setup

We ran all experiments in this paper on a Linux workstation with 2× 3.50 GHz Intel 4-core E5-2637 v2 Xeon CPUs, 256 GB of main memory, and an NVIDIA K40c GPU with 12 GB on-board memory. The GPU programs were compiled with NVIDIA's nvcc compiler (version 8.0.44). The C code was compiled using gcc 4.9.3. All results ignore transfer time (from disk-to-memory and CPU-to-GPU). The merge path operation is from the Modern GPU library [12]. The version of cuSPARSE used was 8.0. The code generated during the current study are available in the figshare repository and GitHub repository [25].

The 157 datasets mentioned in the previous section represent a random sample from the SuiteSparse sparse matrix collection. The topology of the datasets varies from small-degree large-diameter (road network) to scale-free. In the microbenchmark Figure 1(a), dense matrices (varying from 2 rows with 8.3M nonzeroes per row to 8.3M rows with 2 nonzeroes per row) used in the microbenchmark are generated to be nonzero, and converted to CSR sparse matrix storage. We then multiply the matrix by a dense vector and a dense matrix with 64 columns using the vendor-supplied SpMV and SpMM implementations respectively.

### 5.2 Algorithm I: Row-split

Figure 5(a) shows the performance of our row split implementation on 10 SuiteSparse datasets with long matrix rows (62.5 nonzeroes per row on average). We obtain a geomean speed-up of 30.8% over the next fastest implementation and 39% peak improvement.

We suspect our performance drop to the left in Figure 4 comes from the sensitivity to parameter $L$ on row lengths that are significantly less than 32. This causes divergence and uncoalesced memory accesses. On the right hand side, we do much better than cuSPARSE. We believe this is due to the additional occupancy that we can get from superior ILP, which is better at hiding latency. Using the profiler, we noted a 102% improvement in executed instructions per cycle for the matrix sized 128-by-131072.

[3] `https://doi.org/10.6084/m9.figshare.6378764`
[3] `https://github.com/owensgroup/merge-spmm`

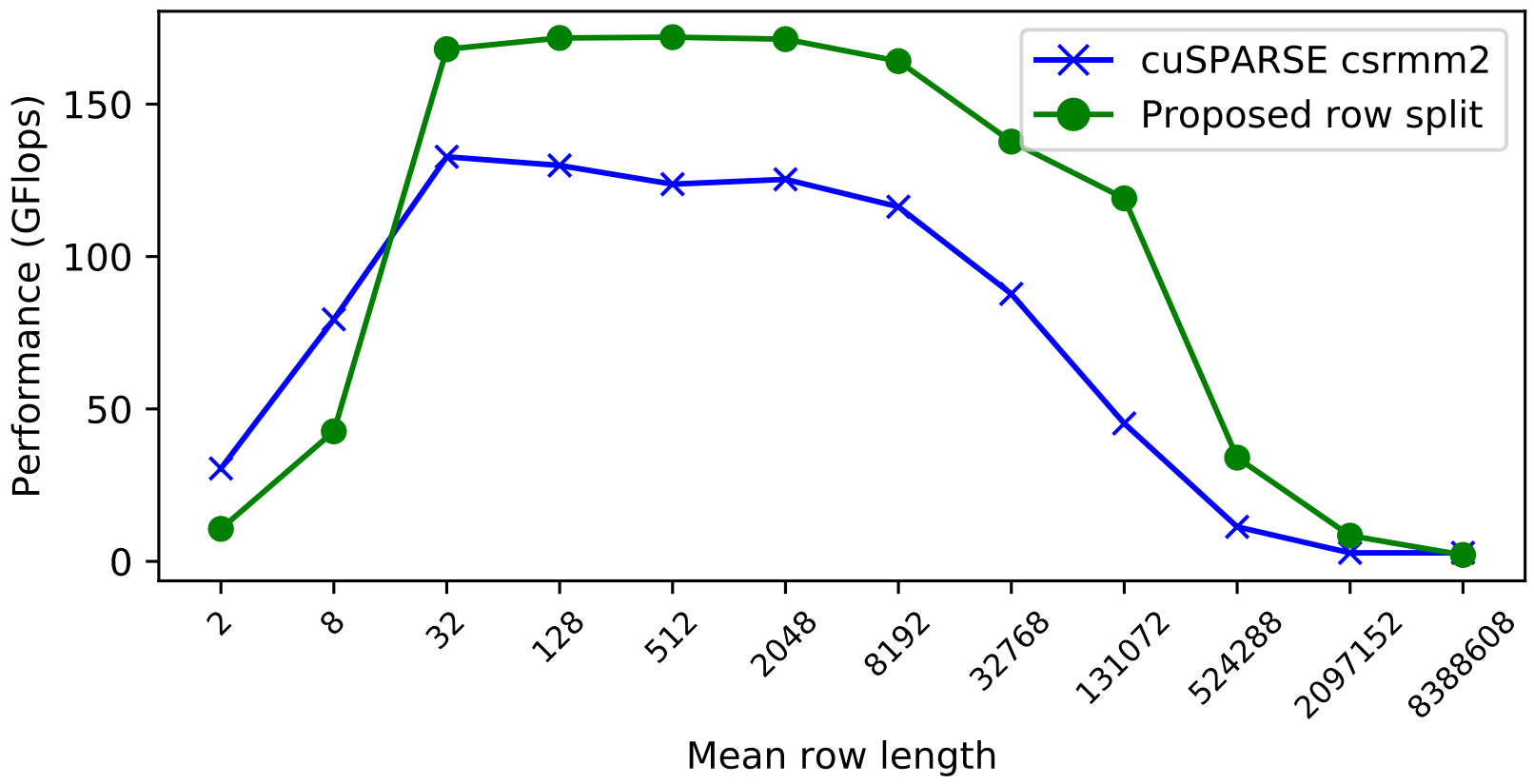

**Fig. 4.** The performance of our proposed SpMM row split kernel vs. NVIDIA cuSPARSE's SpMM as a function of aspect ratio on a Tesla K40c.

We also tried loading in the transpose configuration, where each thread performs a texture load, and loads a different row of the dense matrix. Then, the threads could perform a shuffle reduce, which is a common pattern in GPU programming. However, we observed that this resulted in poorer performance than the vendor-supplied library on average. We suspect the reason for this is there was too much contention amongst different threads for the very limited texture cache resource.

We tried variants that generate the output in column-major order, because this is what cuSPARSE csrmm and csrmm2 produces as output. However, we found that doing such a transpose in the write to global memory causes at most a loss of 3-4 GFlops in performance. The results track Figure 5(a) very closely. Another reason for our performance improvement comes from our use of the shuffle broadcast technique, where we have all 32 threads take turns in broadcasting their values to other threads. This saved shared memory (both in capacity and throughput) which could be put to use elsewhere.

### 5.3 Algorithm II: Merge-based

Figure 5(b) shows the performance of our merge-based SpMM kernel on 10 SuiteSparse datasets with short matrix rows (7.92 nonzeroes on average). We obtain a geomean speed-up of 53% over cuSPARSE csrmm2 and 237% peak improvement. We think the biggest reason that merge path is doing better than the other methods is because it handles Type 2 load imbalances much better. Other methods inevitably encounter warp efficiency degradation due to the divergence

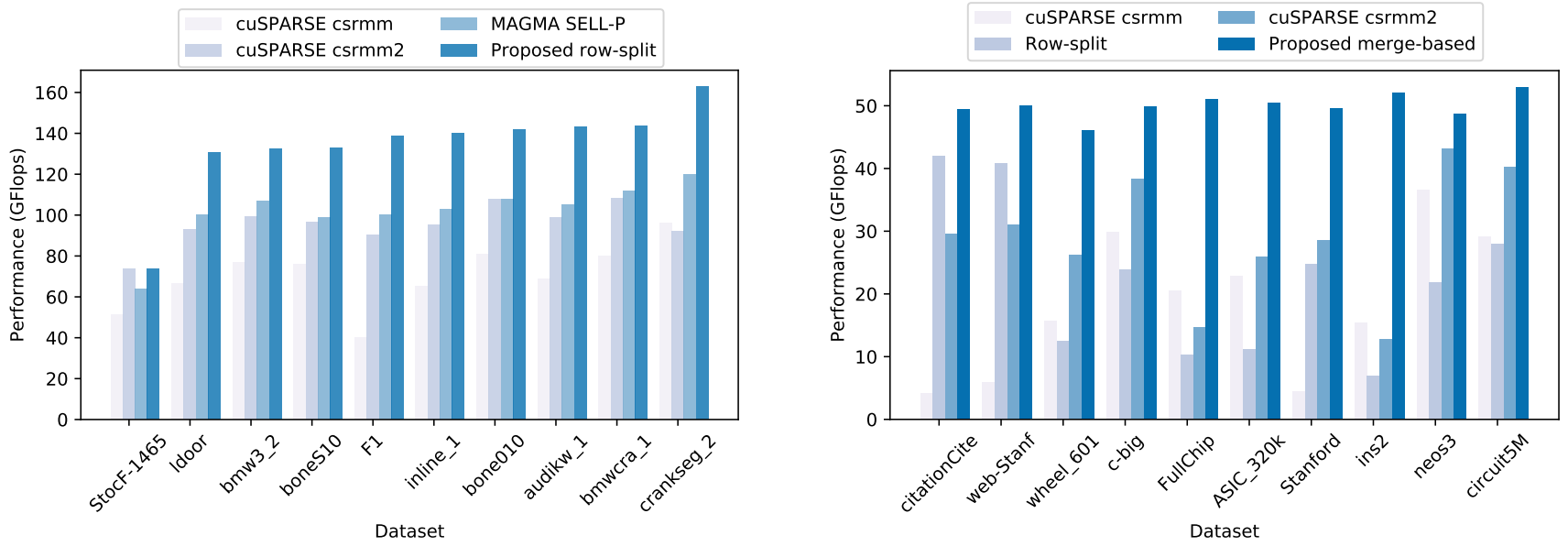

(a) Long rows (62.5 nonzeroes/row) (b) Short rows (7.92 nonzeroes/row)

**Fig. 5.** Performance comparison between the proposed ILP-centric row split kernel and other state-of-the-art kernels on matrices with *long* and *short* row lengths on Tesla K40c using single-precision floating-point. cuSPARSE csrmm and csrmm2 are from a vendor-supplied library [20]. MAGMA SELL-P is by Anzt, Tomov, and Dongarra [17].

caused by short rows, as shown in Figure 1(b). However, merge path can handle these short rows very well by simply allocating more rows to a CTA if the rows are short.

Another interesting observation to make is that the merge path performance in Figure 5(b) all tend to be lower than their row split equivalents. This means that merge path has more overhead than row split, so it is only worth it to perfectly load-balance matrices when it is profitable to do so (Section 5.4). While Merrill and Garland found their merge-based solution was better than row split on SpMV [14], ours did not perform as well on SpMM, as explained in the next paragraph.

As Table 1 shows, merge path's advantage in SpMV comes from being able to obtain $T$ times more ILP per thread than row split, but it enjoys no such advantage in SpMM, where row splitting gets as much ILP as there are nonzeroes in the sparse matrix row as long as row split can afford to pay the register cost. This can be seen in Figure 3(a). While merge path has the opportunity to obtain $T$ times more ILP, we discovered that we need to keep $T = 1$ in order to keep the register count manageable. In typical merge path SpMV implementations, $T$ can be as high as 7. The ILP advantage merge-based had in SpMV is not so assured.

### 5.4 Heuristic

By comparing the speed-up of row split and merge-based to the fastest vendor-supplied SpMM on 157 SuiteSparse sparse matrix collection datasets [15] (see Figure 6(a)), we show that the two proposed algorithms achieve speed-ups over the SpMM state-of-the-art in separate regions on the spectrum of matrix irregularity. However, the geomean speed-up is only a 13.2% gain and 21.5% slowdown for row split and merge-based respectively.

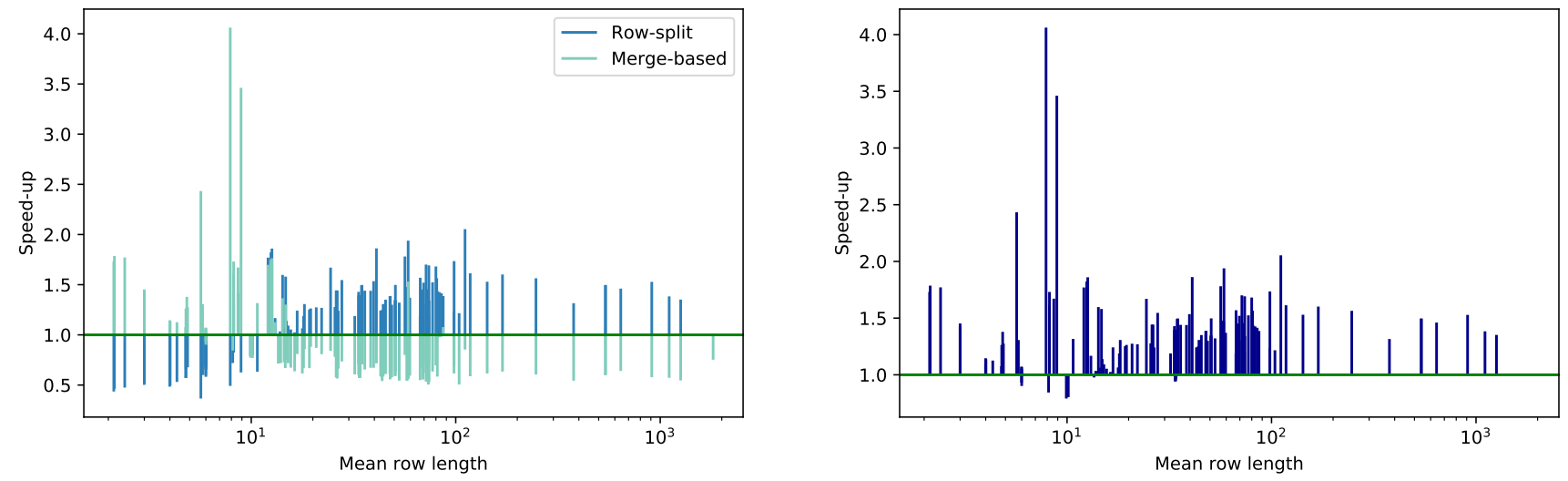

(a) Row split and merge-based separately vs. cuSPARSE csrmm2.

(b) Combined row split and merge-based vs. cuSPARSE csrmm2.

**Fig. 6.** Performance comparison between proposed row split kernel, proposed merge-based kernel, and cuSPARSE csrmm2 on 195 non-trivial datasets from the SuiteSparse sparse matrix collection [15].

Therefore, we propose a heuristic for switching between them using an inexpensive $O(1)$ calculation $d = \frac{nnz}{n}$. Our heuristic is simply computing the average row length for the matrix, and using this value to decide whether to use merge-based or row split. To pinpoint the transition point, we examine Figure 6(a). For our heuristic, we decide that we will use merge-based on datasets whose mean row length is less than 9.35, and row split otherwise.

Using this heuristic, we obtain an overall 31.7% geomean speed-up, and up to a peak of 4.1×, over the vendor-supplied library cuSPARSE csrmm2. Over cuSPARSE csrmm, we obtain a 2.69× geomean speed-up and 22.4× peak speed-up. The result is shown in Figure 6. Using this heuristic as a binary classifier, we get 99.3% accuracy vs. an oracle that perfectly chooses the fastest implementation.

## 6 Conclusion and Future Work

In this paper we implement two promising algorithms for computing sparse matrix dense matrix multiplication on the GPU. Our results using SpMM show considerable performance improvement over the vendor-supplied SpMM on a wide spectrum of graphs. One of the keys to our high performance is our memory-access strategy that allows coalesced access into all 3 matrices (see Figure 3(a)).

In Figure 7, we generate a $100,000 \times 100,000$ random matrix by making a fixed percentage of elements in each row nonzero by sampling indices between 1 and 100,000 without replacement. Our experiments indicate that when multiplying a sparse matrix randomly generated thusly with a tall-skinny dense matrix of size $100,000 \times 64$, our proposed merge-based SpMM is faster than a dense matrix-dense matrix (GEMM) multiplication when less than 9% of the sparse matrix is filled.

Greiner and Jacob have proven theoretically [26] that as the number of nonzeroes per row exceeds some hardware threshold, namely $\frac{m}{M}$ where $m$ is the number of rows in the sparse matrix and $M$ is the size of the fast memory of the device, tiling will become more efficient than the access pattern described in this paper (i.e. going across the sparse matrix and selecting nonzeroes in the dense matrix). Indeed, they claim that tiling both the sparse matrix $\mathbf{A}$ and $\mathbf{B}$ in a manner akin to tiling dense matrix-matrix multiplication is optimal. In future work, it would be interesting to find out whether doing this tiling will extend SpMM's effectiveness range beyond 9% sparsity.

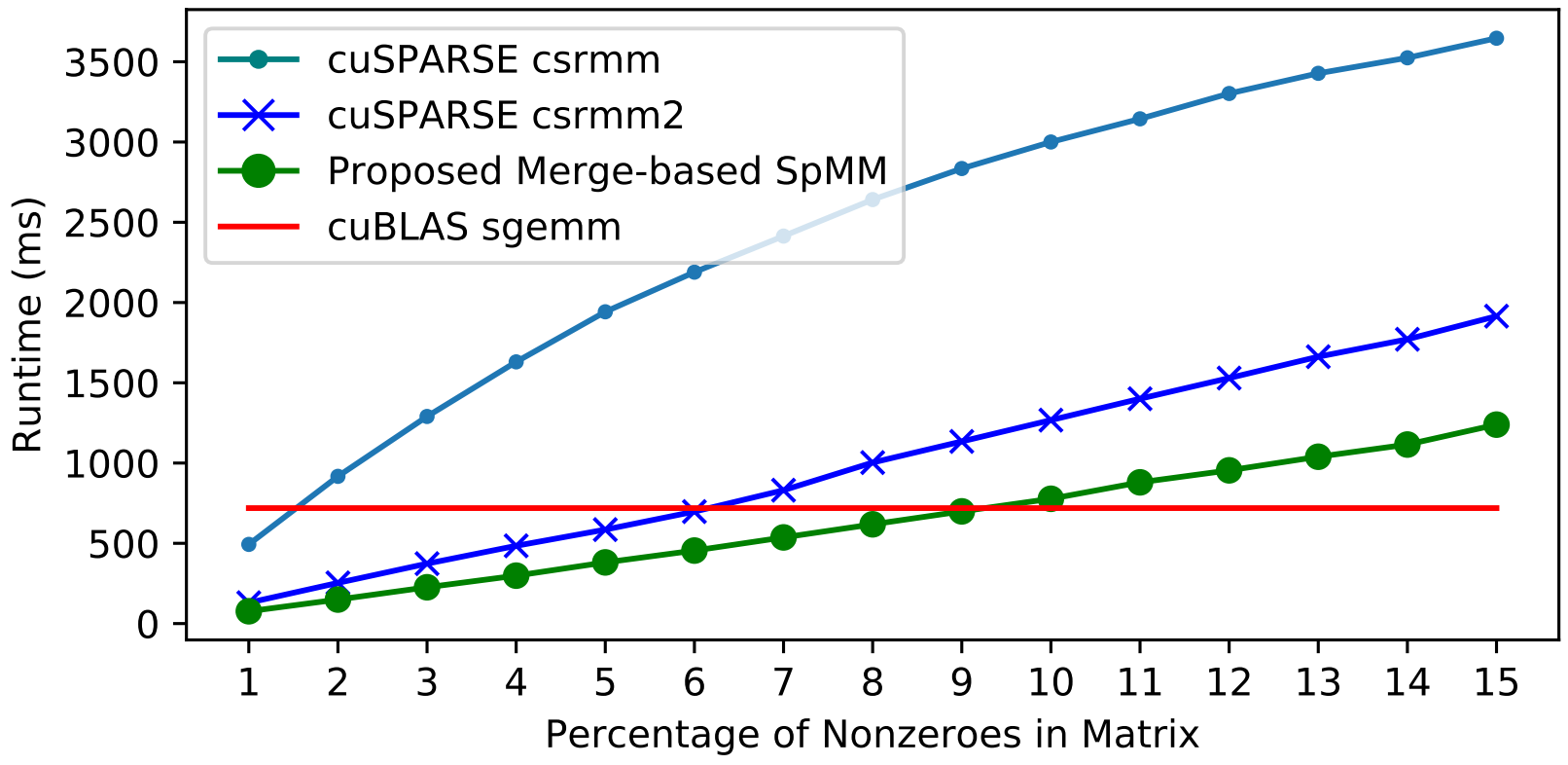

**Fig. 7.** Runtime as a function of the percentage of nonzeroes in the sparse matrix on Tesla K40c using single-precision floating-point. cuSPARSE csrmm, csrmm2 are sparse-dense matrix multipilcation functions from a vendor-supplied library [20]. cuBLAS sgemm is a dense-dense matrix multiplication function from a vendor-shipped library.

Our codes only use the popular CSR data structure, hence avoiding the penalty of sparse matrix format conversions. There are legitimate reasons for considering other formats. For example, certain iterative algorithms require multiplication of a sparse matrix (SpMM) as well as its transpose (SpMM_T) within the same code. Compressed Sparse Blocks (CSB) [27] is a format that is specifically designed for this task and it has already been utilized for SpMM and SpMM_T [18] in CPUs. However, achieving high performance with CSB on irregular matrices requires an efficient load balancer and it is not clear whether GPUs are suitable for this task.

An interesting future direction for research is designing a library around load-balancing techniques such as merge path. While merge path is already present in two libraries–Modern GPU and CUB [12, 28]—they are not designed as layers separated from computation. Similarly in our code, computation and load-balancing are very tightly knit. It would be interesting to discover how to abstract out the load balancing from the computation. Ideally, the user would have to identify the quantities that are desirable for load balancing separately from the computation. Then the load-balancing library would handle the rest making load-balanced GPU kernels much easier to write. The impact of our improved SpMM kernels on application codes is also worth investigating in the future. In particular, we expect a co-design approach to provide more pronounced performance benefits to applications compared to drop-down kernel replacement.

## Acknowledgments

We appreciate the funding support from the National Science Foundation (Award # CCF-1629657), the DARPA XDATA program (US Army award W911QX-12-C-0059), and the DARPA HIVE program. For HIVE support, this material is based on research sponsored by Air Force Research Lab (AFRL) and the Defense Advanced Research Projects Agency (DARPA) under agreement number FA8650-18-2-7836. The U.S. Government is authorized to reproduce and distribute reprints for Governmental purposes notwithstanding any copyright notation thereon. The views and conclusions contained herein are those of the authors and should not be interpreted as necessarily representing the official policies or endorsements, either expressed or implied, of Air Force Research Lab (AFRL) and the Defense Advanced Research Projects Agency (DARPA) or the U.S. Government.

This manuscript has been authored by an author at Lawrence Berkeley National Laboratory under Contract No. DE-AC02-05CH11231 with the U.S. Department of Energy. The U.S. Government retains, and the publisher, by accepting the article for publication, acknowledges, that the U.S. Government retains a non-exclusive, paid-up, irrevocable, world-wide license to publish or reproduce the published form of this manuscript, or allow others to do so, for U.S. Government purposes.

This research was supported in part by the Applied Mathematics program of the DOE Office of Advanced Scientific Computing Research under Contract No. DE-AC02-05CH11231, and in part the Exascale Computing Project (17-SC-20-SC), a collaborative effort of the U.S. Department of Energy Office of Science and the National Nuclear Security Administration.